# Ambulatory blood pressure monitoring versus office blood pressure measurement: Are there sex differences?

Aleksandar Miladinović[a,b,*], Miloš Ajčević[b], Giulia Siveri[b], Laura Liguori[b], Lorenzo Pascazio[c], Agostino Accardo[b]

[a]*Institute for Maternal and Child Health – IRCCS, Burlo Garofolo, Trieste, 34137, Italy*
[b]*Department of Engineering and Architecture at the University of Trieste, Trieste, 34127, Italy*
[c]*Department of Medicine, Surgery and Health Sciences at the University of Trieste, Trieste, 34149 Italy*

**Abstract**

The accurate measurement of blood pressure (BP) is an important prerequisite for the reliable diagnosis and efficient management of hypertension and other medical conditions. Office Blood Pressure Measurement (OBP) is a technique performed *in-office* with the sphygmomanometer, while Ambulatory Blood Pressure Monitoring (ABPM) is a technique that measures blood pressure during 24h. The BP fluctuations also depend on other factors such as physical activity, temperature, mood, age, sex, any pathologies, a hormonal activity that may intrinsically influence the differences between OBP and ABPM. The aim of this study is to examine the possible influence of sex on the discrepancies between OBP and ABPM in 872 subjects with known or suspected hypertension. A significant correlation was observed between OBP and ABPM mean values calculated during the day, night and 24h (ABPMday, ABPMnight, ABPM24h) in both groups (p<0.0001). The main finding of this study is that no difference between sexes was observed in the relation between OBP and mean ABMP values except between systolic OBP and systolic ABPM during the night. In addition, this study showed a moderate correlation between BPs obtained with the two approaches with a great dispersion around the regression line which suggests that the two approaches cannot be used interchangeably.

*Keywords:* blood pressure; ambulatory blood pressure monitoring; ABPM;

---

\* Corresponding author. Tel.: +39 040 558 7130
E-mail address: aleksandar.miladinovic@burlo.trieste.it

**1. Introduction**

The accurate measurement of blood pressure (BP) is an important prerequisite for the reliable diagnosis and efficient management of hypertension and other medical conditions. Office Blood Pressure Measurement (OBP) is a technique performed *in-office* with the sphygmomanometer by specifically trained clinical staff and it is considered the standard technique for measuring BP for a diagnosis of hypertension. On the other hand, Ambulatory Blood Pressure Monitoring (ABPM) is a technique that measures blood pressure over 24h. The acquired data consists of a series of BP measurements automatically acquired, at regular intervals, by the instrument during the subject's usual daily activities. The systolic (SBP) and diastolic (DBP) pressure values are typically measured at 15 minutes intervals during the day and every 30 minutes at night. The biggest disadvantage of the ABPM in the past was that this type of measurement usually required expensive devices (i.e. BP holters), which with technological development became more portable and accessible to the general population.

The literature shows that there are discrepancies between OBP and automated measures such as ABPM [1–5] and that former better model circadian changes of BP [6]. From the technical perspective, the difference is due to the fact that ABPM can better estimate the overall BP affected by the circadian rhythm [7], and in a healthy population, it has physiologically lower values during sleep and early in the morning (nocturnal dipping) [8, 9]. Moreover, ABPM helps to identify masked hypertension [4] and white coat hypertension [10] situations. ABPM is also a reliable predictor of clinical outcomes and organ damage [11, 12] especially during the night [12].

The BP fluctuations also depend on other factors such as physical activity, temperature, mood, age, sex, some pathologies, hormonal activity [13], which may intrinsically make OBP less accurate in the BP estimation than ABPM. The incongruence and factors which influence these differences are still debated, however, from the literature it is shown that sex differences in sympathetic neural-hemodynamic balance have implications for human blood pressure regulation [14–17].

Therefore, the aim of this study is to examine the possible influence of sex on the discrepancies between OBP and ABPM in subjects with known or suspected hypertension.

**2. Materials and Methods**

In this study 872 subjects (355 males, 517 females, mean age 64.9 ± 14.2 years) with known or suspected hypertension were enrolled at the Cardiovascular Pathophysiology Unit of Geriatric Clinic of the Trieste University Hospital, Trieste, Italy between July 2016 and September 2016. In this retrospective study, the inclusion criteria were: (1) no clinical evidence of secondary arterial hypertension, (2) absence of clinical evidence of hypertension-related complications, (3) Office BP measures with an SBP value between 70 mmHg and 260 mmHg and a DBP value between 40 mmHg and 150 mmHg. The measurement procedures were in accordance with the institutional guidelines and all the subjects gave their informed consent.

*2.1. Blood Pressure Measurements*

The blood pressure was at first measured in Office condition by trained clinical staff using a standard mercury sphygmomanometer with a cuff of appropriate size. The average of two consecutive measurements was calculated.

Moreover, ambulatory monitoring along the 24-h was carried out by using a Holter Blood Pressure Monitor (Mobil-O-Graph® NG, IEM GmbH Stolberg, Germany), based on an oscillometric technique. The portable monitor was programmed to obtain ABPM on a 15-min interval throughout the daytime (7:00 to 22:00) and each 30-min interval throughout the nighttime (22:00 to 7:00). No patient received additional medication that might affect the circadian blood pressure or heart rate rhythmicity. Participants were given both oral and written instructions on how to use the Holter Blood Pressure Monitor. From the acquired ABPM data average values were calculated during the awake state, from 7:00 to 22:00 (ABPMday), during the nighttime, from 22:00 to 7:00 (ABPMnight), and for the whole 24h period (ABPM24h).

*2.2. Statistical Analysis*

Measured data resulted normally distributed (Lilliefors test). Therefore, the data were presented as means ± SD for each of the measured parameters (OBP, ABPMday, ABPMnight, ABPM24h) and for each group (male and female). The correlations between OBP and ABPM parameters (ABPMday, ABPMnight, ABPM24h) were investigated by the Pearson coefficients of correlation. Regression analysis was performed to quantify the impact of sex on the discrepancies between OBP and mean ABPMs measurements. The differences between obtained regression lines were investigated.

**3. Results**

Clinical and demographic characteristics for each group are reported in Table 1. No difference in age and BP levels were observed between the two groups. In Figure 1. the ABPM data of individual patients plotted against OBP, for each considered period. A significant correlation was observed between OBP and ABPM parameters (ABPMday, ABPMnight, ABPM24h) in both groups (p<0.0001). The slope, intercept of the regression line is reported in Table 2 and plotted respectively in Figure 1. The identified OBP - ABPM relations did not differ between males and females (Table 3), except in the case of the relation between systolic OBP and systolic ABPMnight (p<0.05). There were also no differences in correlation coefficients for investigated relations between two groups; only for the relation between systolic OBP and systolic ABPMnight the correlation in females was higher than the male group (0.4506 vs 0.3066, respectively p =0.025)

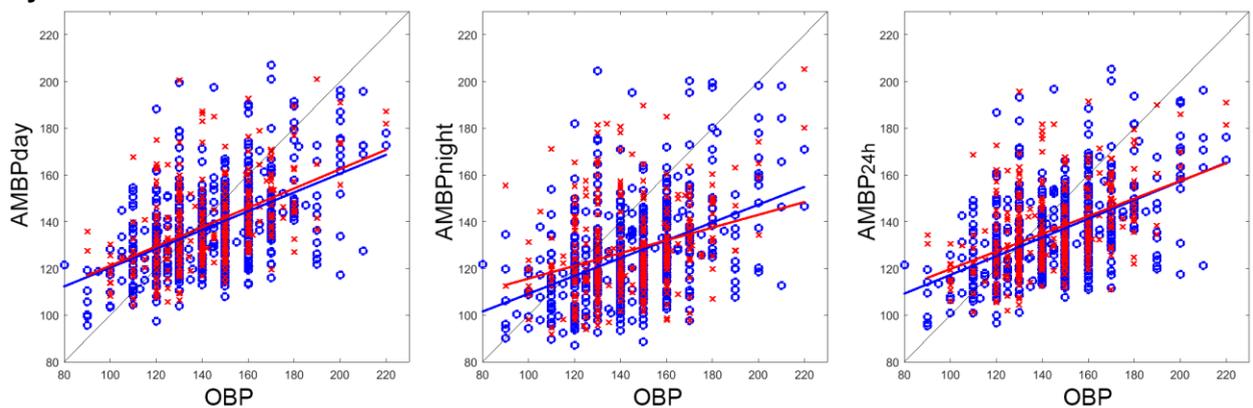

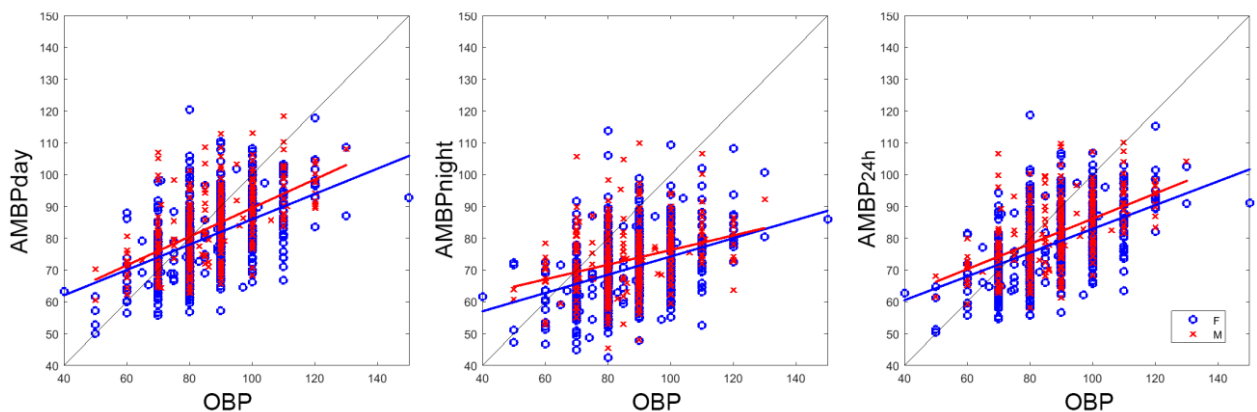

**Figure 1.** Scatter plot with regression lines of mean ABPMs (ABPMday, ABPMnight and ABPM24) data of individual patients plotted against OBP, for systolic (above) and diastolic (bellow) BP.

**Table 1.** Clinical and demographic characteristics females and males

|  |  | *Female (N=517)* | *Male (N=355)* |
|---|---|---|---|
|  | **Age (years)** | 64.3 ± 14.7 | 65.5 ± 14.5 |
| **Systolic** | **OBP (mmHg)** | 145.1 ± 24.2 | 143.1 ± 21.5 |
|  | **ABPM day (mmHg)** | 138.5 ± 19.1 | 138.9 ± 18.0 |
|  | **ABPM night (mmHg)** | 126.3 ± 20.5 | 127.4 ± 19.2 |
|  | **ABPM 24h (mmHg)** | 135.3 ± 18.93 | 136.0 ± 17.5 |
| **Diastolic** | **OBP (mmHg)** | 86.7 ± 14.4 | 86.4 ± 13.2 |
|  | **ABPM day (mmHg)** | 80.6 ± 11.4 | 83.4 ± 10.8 |
|  | **ABPM night (mmHg)** | 70.4 ± 10.7 | 73.1 ± 10.0 |
|  | **ABPM 24h (mmHg)** | 80.6 ± 11.4 | 83.4 ± 10.8 |

\* no significant difference (p<0.05) observed in any of the variables

**Table 2.** The slope, intercept, and Pearson's coefficient r of the regression lines for females and males

|  |  | **FEMALES** | | | **MALES** | | |
|---|---|---|---|---|---|---|---|
|  |  | **slope** | **intercept** | **Pearson's r** | **slope** | **intercept** | **Pearson's r** |
| **Systolic** | **OBP vs ABPMday** | 0.4025 | 80.0801 | 0.5106 | 0.4137 | 79.7613 | 0.4949 |
|  | **OBP vs ABPMnight** | 0.3814 | 71.0273 | 0.4506 | 0.2731 | 88.3301 | 0.3066 |
|  | **OBP vs ABPM24h** | 0.4005 | 77.1834 | 0.5125 | 0.3774 | 81.9848 | 0.4656 |
| **Diastolic** | **OBP vs ABPMday** | 0.3986 | 46.1083 | 0.5041 | 0.4509 | 44.4205 | 0.5478 |
|  | **OBP vs ABPMnight** | 0.2872 | 45.5593 | 0.3859 | 0.2304 | 53.2585 | 0.3025 |
|  | **OBP vs ABPM24h** | 0.374 | 45.5664 | 0.4999 | 0.3969 | 46.4669 | 0.5197 |

**Table 3.** Statistics (z-score and p-values) of the comparison of correlation coefficients and slopes

|  |  | **Z-scores of Pearson's r** | **p-values** | **Z-scores of slopes** | **p-values** |
|---|---|---|---|---|---|
| **Systolic** | **OBP vs ABPMday** | 0.2633 | n.s. | 0.85 | n.s. |
|  | **OBP vs ABPMnight** | 2.2446 | 0.025 | 1.98 | 0.049 |
|  | **OBP vs ABPM24h** | 0.7764 | n.s. | 0.64 | n.s. |
| **Diastolic** | **OBP vs ABPMday** | -0.7464 | n.s. | 0.28 | n.s. |
|  | **OBP vs ABPMnight** | 1.2826 | n.s. | 0.25 | n.s. |
|  | **OBP vs ABPM24h** | -0.3336 | n.s. | 0.62 | n.s. |

\* *n.s. – non-significant*

**4. Discussion**

The OBP measurement has been traditionally used as the standard technique for the diagnosis of hypertension [18]. However, in a research setting the OBP can be relatively inaccurate [19, 20]. Several studies suggest that clinical BP assessments performed with OBP do not correctly diagnose hypertension due to the white coat effect and masked hypertension [19, 20]. The European Society of Hypertension/European Society of Cardiology and American College of Cardiology/American Heart Association hypertension guidelines [13, 21] acknowledged the importance of new and fully automated techniques, such as ABPM. Several factors have been shown to affect the occurrence of the white coat effect, such as age, sex, and nonsmoking habit [13, 22]. Moreover, the white coat effect produces OBP values higher than ABPM, as shown in [23]. The differences in BPs measured with these two techniques in the general population were reported in recent studies [1–5, 24, 25]. In particular, ABPM is a much more reliable predictor of clinical outcomes and organ damage [11, 12, 26]; in this last case, the night ABPM measurement is considered more informative [12]. Therefore, the relationship between BP measurements performed in the office and calculated during 24h, day and night have drawn the attention of the scientific community [18, 27] and the causes of the differences between their values are still debated. Therefore, we aimed at evaluating a possible confounding effect of sex on the observed difference between OBP and mean ABMPs (ABPMday, ABPMnight, ABPM24h).

The main finding of this study is that no difference between sexes was observed in the relation between OBP and mean ABMP values except between systolic OBP and systolic ABPM during the night, where the correlation was weak for the male group. In addition, this study showed a moderate correlation between BPs obtained with two approaches. Indeed, although the correlation was significant there was a great dispersion around the regression line which suggests that the two approaches cannot be used interchangeably. Furthermore, we observed that OBP underestimated ABMP values in the lower pressure range and vice versa overestimates for the higher pressure range which is in line with the study [28]. Further studies are needed to identify the confounding factors.

**Acknowledgements**. Work partially supported by the Master's program in Clinical Engineering, University of Trieste.

**Conflict of Interest.** The authors declare that the research was conducted in the absence of any commercial or financial relationships that could be construed as a potential conflict of interest.